\documentclass[11pt]{article}
\usepackage{graphicx}
\usepackage{subfloat}
\usepackage{epsfig}
\usepackage{amsfonts}   
\usepackage{amssymb}

\begin{document}
\date{}
\title{{\bf{\Large Effect of external magnetic field on holographic superconductors in presence of nonlinear corrections}}}
\author{
 {\bf {\normalsize Dibakar Roychowdhury}$
$\thanks{E-mail: dibakar@bose.res.in, dibakarphys@gmail.com}}\\
 {\normalsize S.~N.~Bose National Centre for Basic Sciences,}
\\{\normalsize JD Block, Sector III, Salt Lake, Kolkata-700098, India}
\\[0.3cm]
}

\maketitle
\begin{abstract}
 In this paper, based on analytic technique, several properties of holographic $ s $-wave superconductors have been investigated in the presence of various higher derivative (\textit{non linear}) corrections to the usual Maxwell action. Explicit expressions for the critical temperature and the condensation values have been obtained in the probe limit. Finally, the nature of condensate solutions have been investigated by immersing the superconductor in an external magnetic field.  It is found that below certain critical magnetic field strength ($ B_c $) there exists a superconducting phase. Most importantly it has been observed that the value of this critical field strength ($ B_c $) indeed gets affected due to the presence of higher derivative corrections to the usual Maxwell action. 
\end{abstract}
\section{Introduction}
 The BCS theory \cite{ref1} of superconductivity has been the most successful microscopic theory to describe various properties of usual (low temperature) superconducting materials (including many metallic elements like, Pb, Al etc.) with great accuracy. On the other hand there are another class of superconductors, that are supposed to be strongly coupled, where the understanding of pairing mechanism  responsible for superconductivity remains completely obscure. Interestingly one can attempt to answer these questions using the AdS/CFT correspondence \cite{ref2}-\cite{ref5}, where one is tempted to exploit the gauge/gravity duality \cite{ref6} in order to study various properties of strongly coupled phenomena in usual QFT that also arise in various condensed matter systems.
 
It was Gubser \cite{ref7}-\cite{ref9}, who first argued that the gravity dual of a superconductor could be found through the mechanism of spontaneous $ U(1) $ symmetry breaking near the black hole event horizon which results in a condensation of scalar hair at a temperature ($ T $) that is less than certain critical value ($ T_c $). This critical value ($ T_c $) below which the scalar hair forms may be identified as the critical temperature corresponding to a second order phase transition from a normal phase to a superconducting phase in the dual field theory. It is in fact the local $ U(1) $ symmetry breaking in the bulk which corresponds to a global $ U(1) $ symmetry breaking in the dual field theory residing at the boundary of the AdS space and thereby inducing a superconductivity.  Later on this idea was further developed and systematically extended by Horowitz et al \cite{ref10}-\cite{ref13}, who have found that such a simple gravitational dual can indeed reproduce all the standard features of the conventional superconductors \cite{ref14}. From their analysis it is also evident that these superconductors actually mimic high $ T_c $ superconductors in various respects. For excellent reviews see \cite{ref15}-\cite{ref17}. Following this remarkable correspondence, till date a number of investigations have been performed in various directions \cite{ref18}-\cite{nref3}.

One of the major characteristic properties of ordinary superconductors is that they exhibit perfect diamagnetism as the temperature is lowered through $ T_c $ in presence of an external magnetic field. In other words, at low temperature superconductors expel magnetic field lines. This is known as \textit{Meissner} effect which could be put into the form of a following parabolic law \cite{ref14},
\begin{equation}
B_c(T)\approx B_c(0)\left[1-\left(\frac{T}{T_c} \right)^{2}  \right]. 
\end{equation}
 Depending on their behavior in the presence of an external magnetic field, ordinary superconductors are classified into two categories, namely type I and type II. In type I superconductors, for $ B>B_c $ there exists a first order phase transition from the superconducting phase to the normal phase where $ B_c $ is the value of the critical field strength. Whereas, on the other hand, in type II superconductors there happens to be a gradual second order phase transition and the material ceases to super conduct for $ B>B_{c2} $ where $ B_{c2} $ is the upper critical field strength.

Inspired from all these facts, till date a number of attempts have been made in order to investigate the effects of applying an external magnetic field to  holographic superconductors \cite{ref30}-\cite{ref34}. From these analysis, it is more or less confirmed that holographic superconductors are of type II rather than type I, which is also in agreement with their so called high $ T_c $ behavior. In spite of all these attempts, some crucial issues are yet to be explored, which may be put as follows:
\vskip 1mm
\noindent
$\bullet$  Since all the above attempts are mostly concerned with \textit{ numerical} techniques, the question that naturally arises is that whether it is possible, in general, to have an \textit{analytic} scheme which could be employed to investigate the behavior of holographic superconductors even in the presence of an external magnetic field. Furthermore, it is also to be noted that all the above analysis are mostly performed for ($ 2+1 $) dimensional holographic superconductors. Therefore, it will be quite interesting to see whether one can also carry out analytic calculations for ($ 3+1 $) dimensional case. In other words, whether holographic superconductors are still type II in ($ 3+1 $) dimensions. 
\vskip 1mm
\noindent
$\bullet$ Most importantly, one may note that, so far all the attempts to study the effect of magnetic field on a holographic superconductor are made in the framework of usual Maxwell electrodynamics. Therefore, it will be quite natural to ask how the higher derivative corrections of the gauge fields can affect the behavior of holographic superconductors in the presence of an external magnetic field. These higher derivative corrections could be incorporated in the theory of superconductors replacing the Maxwell action in \cite{ref8} by a non linear action for classical electrodynamics. 
 
In order to address the above mentioned issues, in the present work, based on an \textit{analytic} scheme \cite{ref19}, we investigate the effect of adding an external magnetic field on holographic $ s $- wave condensate in the presence of both (i) Born-Infeld (BI) \cite{ref35} as well as the (ii) Weyl curvature corrections \cite{nref4} to the usual Maxwell action. Considering the probe limit the entire analysis has been carried out in the back ground of a planar Schwarzschild AdS space time. At this stage it is reassuring to note that these non linear generalizations essentially correspond to the higher derivative corrections of the gauge fields. In both the cases the analytic expressions for the critical temperature ($ T_c $) as well as the condensation values have been obtained, where the computations are performed in the leading order in the coupling  parameter(s). It has been observed that the coupling parameter(s) of the theory indeed affect the formation of scalar hair at low temperatures. Finally, the effect of an external magnetic field on the formation of holographic $ s $- wave condensate has been studied by adding a static magnetic field in the bulk theory. Interestingly, we find that the superconducting phase disappears for $ B>B_c $, where $ B_c $ is the critical field strength. Moreover, we note that the value of this critical field strength ($ B_{c} $) indeed gets affected due to the presence of non linearity in the original theory, which also in turn affects the formation of scalar hair at low temperatures.

Before going further, let us briefly mention about the organization of the paper. In section 2, we investigate the effect of external magnetic field on the $ s $- wave holographic superconductors in the presence of BI corrections to the usual Maxwell action.  In section 3, similar analytic computations have been carried out incorporating the effect of Weyl curvature corrections in the original Maxwell theory.  Finally, the paper is concluded in section 4.

\section{Magnetic field effect with Born-Infeld corrections}

In the present section, considering the probe limit, we aim to discuss the effect of an external magnetic field on the holographic $ s $- wave condensate in the presence of Born-Infeld (BI) corrections to the usual Maxwell action.
We begin with the metric of a planar Schwarzschild AdS black hole, which may be written as \cite{ref10},
\begin{eqnarray}
ds^2=-f(r)dt^2+\frac{1}{f(r)}dr^2+r^2(dx^2+dy^2)
\label{m1}
\end{eqnarray}
with
\begin{eqnarray}
f(r)=r^{2}-\frac{r_{+}^3}{r}
\label{metric}
\end{eqnarray}
in units in which the AdS radius is unity, i.e. $l=1$.
The Hawking temperature is related to the horizon radius ($r_+$) as
\begin{eqnarray}
T=\frac{3r_+}{4\pi}~.
\label{ntemp}
\end{eqnarray}
The entire analysis have been performed over this fixed back ground.

In order to study the holographic dual of this theory we adopt the following action which includes a complex scalar field minimally coupled to the Maxwell field $ A_{\mu} $ as,
\begin{equation}
S= \frac{1}{16\pi G_4}\int d^{4}x\sqrt{-g}\left[R-2\Lambda +\frac{1}{b}\bigg(1-\sqrt{1+\frac{b F}{2}}\bigg)-|\nabla_{\mu}\psi-iA_{\mu}\psi|^2-m^2|\psi|^2 \right], 
\end{equation}  
where, $ F= F_{\mu\nu}F^{\mu\nu}$, $ b $ is the BI coupling parameter and $ \Lambda\left( =-\frac{3}{l^{2}}\right)  $ is the cosmological constant. It is reassuring to note that in the limit $ b\rightarrow 0 $ one recovers the usual Maxwell action. It is to be noted that the higher order terms in the coupling parameter $ b $ essentially implies the higher derivative corrections of the gauge fields. 

In the probe limit, the Maxwell and scalar field equations may be found as,
\begin{eqnarray}
\frac{1}{\sqrt{-g}}\partial_{\mu}\left( \frac{\sqrt{-g}F^{\mu\nu}}{\sqrt{1+\frac{bF}{2}}}\right)-i\left( \psi^{*}\partial^{\nu}\psi - \psi(\partial^{\nu}\psi)^{*}\right)-2A^{\nu}|\psi|^{2} = 0 \label{eq1}
\end{eqnarray} 
and,
\begin{equation}
\partial_{\mu}\left(\sqrt{-g}\partial^{\mu}\psi \right) -i\sqrt{-g}A^{\mu}\partial_{\mu}\psi -i\partial_{\mu}\left(\sqrt{-g}A^{\mu}\psi \right)-\sqrt{-g}A^{2}\psi -\sqrt{-g}m^{2}\psi =0\label{eq2} 
\end{equation}
respectively.

In order to solve the the above set of equations(\ref{eq1},\ref{eq2}) we adopt the following ansatz \cite{ref10},
\begin{eqnarray}
A_{\mu}=(\phi(r),0,0,0),\;\;\;\;\psi=\psi(r)
\label{vector}
\end{eqnarray}
which finally yields the following set of equations,
\begin{eqnarray}
\phi''(r)+\frac{2}{r}\phi'(r)\bigg(1-b\phi'^2 (r)\bigg)
-\frac{2\psi^2 (r)}{f}\phi(r)\bigg(1-b\phi'^2 (r)\bigg)^{3/2}=0\label{neq3}
\end{eqnarray}
and,
\begin{eqnarray}
\psi^{''}(r)+\left(\frac{f'}{f}+\frac{2}{r}\right)\psi'(r)
+\left(\frac{\phi^{2}(r)}{f^2}-\frac{m^2}{f}\right)\psi(r)=0\label{neq4}
\end{eqnarray}
respectively.\\

Setting $ m^{2}=-2 $, and changing the variable from $ r $ to $ z(=\frac{r_+}{r}) $ the above set of equations (\ref{neq3},\ref{neq4}) turn out to be,
\begin{eqnarray}
\phi''(z)+\frac{2bz^3}{r_{+}^2 }\phi'^{3}(z)-\frac{2\psi^2 (z)\phi(z)r_+^{2}}{z^4 f(z)}
\left(1-\frac{bz^4}{r_{+}^2}\phi'^{2}(z)\right)^{3/2} =0\label{neq5}
\end{eqnarray}
and,
\begin{eqnarray}
\psi''(z)+\frac{f^{'}(z)}{f(z)}\psi'(z)+ \frac{r_+^{2}\phi^{2}(z)\psi(z)}{z^{4}f^{2}(z)}+\frac{2r_+^{2}\psi(z)}{z^{4}f(z)}=0\label{neq6}
\end{eqnarray}
respectively.

Let us now talk about the boundary conditions:
\vskip 1mm
\noindent
$\bullet$ Regularity at the horizon $ z=1 $ implies,
\begin{eqnarray}
\phi (1)=0,~~~\psi^{'}(1)=\frac{2}{3}\psi(1) .\label{neq7}
\end{eqnarray}

\vskip 1mm
\noindent
$\bullet$ In the asymptotic region ($ z\rightarrow 0 $) the solutions may be written as,
\begin{eqnarray}
\phi(z) = \mu - \frac{\rho}{r_+}z,~~~~\psi(z)= J_{-}z + J_{+}z^{2},\label{neq8}
\end{eqnarray}  
where $ \mu $ and $ \rho $ are the chemical potential and the charge density of the dual field theory. In the following analysis we set $ J_{-}=0 $ .

With the above expressions in hand, as a next step, we aim to derive an analytic expression for the critical temperature and the condensation values in the presence of above non linear (BI) corrections to the usual Maxwell action. In order to do that, we first Taylor expand both $ \phi(z) $ and $ \psi(z) $ near the horizon as,
\begin{eqnarray}
\phi (z)=\phi (1) - \phi^{'}(1)(1-z) + \frac{1}{2}\phi^{''}(1)(1-z)^{2} + .. ..\label{neq9}
\end{eqnarray}
and,
\begin{eqnarray}
\psi(z)=\psi(1)-\psi^{'}(1)(1-z)+\frac{1}{2} \psi^{''}(1)(1-z)^{2} + ....\label{neq10}
\end{eqnarray}
respectively, where without loss of generality we choose $ \phi'(1)<0 $ and $ \psi(1)>0 $.

On the other hand, near $ z=1 $ from (\ref{neq5}) we obtain,
\begin{equation}
\phi^{''}(1)= -\frac{2b \phi'^{3}(1)}{r_+^{2}}-\frac{2\psi^2 (1)\phi'(1)}{3}\left( 1-\frac{3b}{2r_+^{2}}\phi'^{2}(1)\right) 
+ \mathcal{O}(b^{2})\label{neq11},
\end{equation}
where we have used the fact that near the event horizon ($ z=1 $) the function $ f(z) $ could be Taylor expanded as in (\ref{neq9},\ref{neq10}). Finally, substituting (\ref{neq11}) into (\ref{neq9}) we obtain,
\begin{equation}
\phi(z)= -\phi'(1)(1-z)-\left[\frac{b \phi'^{2}(1)}{r_+^{2}}+\frac{\psi^{2}(1)}{3}\left( 1-\frac{3b}{2r_+^{2}}\phi'^{2}(1)\right)  \right] \phi'(1)(1-z)^{2} + \mathcal{O}(b^{2})\label{neq12}.
\end{equation}
Similarly, from (\ref{neq6}) and using (\ref{neq7}), near $ z=1 $ we obtain,
\begin{equation}
\psi''(1)= \frac{8}{9}\psi(1) -\frac{\psi(1)\phi'^{2}(1)}{18r_+^{2}}\label{neq13}.
\end{equation} 
Substituting (\ref{neq13}) into (\ref{neq10}) we finally obtain,
\begin{equation}
\psi (z) = \frac{1}{3}\psi(1)+\frac{2}{3}\psi(1)z+\left( \frac{4}{9}-\frac{\phi'^{2}(1)}{36r_+^{2}}\right)\psi(1)(1-z)^{2}\label{neq14}. 
\end{equation}

Following the methodology developed in \cite{ref19}, one can obtain an analytic expression for the critical temperature ($ T_c $) by matching the solutions (\ref{neq8}), (\ref{neq12}) and (\ref{neq14}) at some intermediate point $ z=z_m $.

In order to match these two sets of asymptotic solutions smoothly at $ z=z_m $ we need the following four conditions,
\begin{eqnarray}
\mu -\frac{\rho z_m}{r_+} = \beta(1-z_m) + \beta\left[\frac{b \beta^{2}}{r_+^{2}} +\frac{\alpha^{2}}{3}\left(1-\frac{3b\beta^{2}}{2r_+^{2}} \right)  \right] (1-z_m)^{2}\label{neq15}
\end{eqnarray}
\begin{equation}
-\frac{\rho}{r_+} = -\beta - 2\beta (1-z_m)\left[\frac{b \beta^{2}}{r_+^{2}} +\frac{\alpha^{2}}{3}\left(1-\frac{3b\beta^{2}}{2r_+^{2}} \right)  \right] \label{neq16}
\end{equation}
\begin{eqnarray}
 J_+z_m^{2} = \frac{\alpha}{3}+\frac{2\alpha z_m}{3} + \alpha \left(\frac{4}{9}-\frac{{\tilde{\beta^{2}}}}{36} \right)(1-z_m)^{2}\label{neq17} 
\end{eqnarray}
and,
\begin{equation}
J_+z_m = \frac{\alpha}{3} - \alpha \left(\frac{4}{9}-\frac{{\tilde{\beta^{2}}}}{36} \right)(1-z_m) \label{neq18}
\end{equation}
where we have set $ \beta=-\phi'(1) $, $ \alpha = \psi(1) $ and $ \tilde{\beta}=\frac{\beta}{r_+} $.

As a first step, from (\ref{neq16}) we obtain,
\begin{equation}
\alpha^{2}= \frac{3}{2(1-z_m)}\left[\left(\frac{\rho}{\beta r_+} -1\right)+ \frac{b\beta^{2}}{r_+^{2}} \left(\frac{3\rho}{2\beta r_+}-\frac{(7-4z_m)}{2} \right)  \right] + \mathcal{O}(b^{2})\label{neq19}.
\end{equation}
Using (\ref{ntemp}), from (\ref{neq19}) we finally obtain,
\begin{equation}
\alpha^{2}=\frac{3}{2(1-z_m)} \left(\frac{T_c}{T}\right)^{2}\left(1+\frac{b\tilde{\beta^{2}}}{2}(7-4z_m)\right) \left(1-\frac{T^{2}}{T_c^{2}} \right) + \mathcal{O}(b^{2})\label{neq20},  
\end{equation}
where,
\begin{equation}
T_c = \frac{3\sqrt{\rho}}{4\pi\sqrt{\tilde{\beta}}}\sqrt{1-2b\tilde{\beta^{2}}(1-z_m)}\label{neq21}.
\end{equation}
Finally, for $ T\sim T_c $ i.e, very close to the critical temperature, from (\ref{neq20}) we obtain,
\begin{equation}
\alpha = \psi(1) = \sqrt{\frac{3}{1-z_m}} \left(1+\frac{b\tilde{\beta^{2}}}{4}(7-4z_m)\right) \sqrt{1-\frac{T}{T_c}}+\mathcal{O}(b^{2})\label{neq22}.
\end{equation}

From (\ref{neq17}) and (\ref{neq18}) one can further obtain,
\begin{equation}
J_+ = \frac{\alpha(2+z_m)}{3z_m} ~~ and,~~~ \tilde{\beta}= 2\sqrt{\frac{7-z_m}{1-z_m}}\label{neq23}.
\end{equation}

Finally, using (\ref{ntemp}), (\ref{neq22}) and (\ref{neq23}), near the critical temperature ($ T\sim T_c $) the condensation operator may be calculated as \cite{ref19},
\begin{eqnarray}
<\mathcal{O}_2> &=& \sqrt{2} J_+ r_+^{2}\nonumber\\
&=& \frac{16\sqrt{2} \pi^{2}}{9}\left(\frac{2+z_m}{3z_m} \right)\sqrt{\frac{3}{1-z_m}} T_c^{2}  \left(1+\frac{b\tilde{\beta^{2}}}{4}(7-4z_m)\right) \sqrt{1-\frac{T}{T_c}}+\mathcal{O}(b^{2}).
\end{eqnarray}

\begin{figure}[h]
\centering
\includegraphics[angle=0,width=16cm,keepaspectratio]{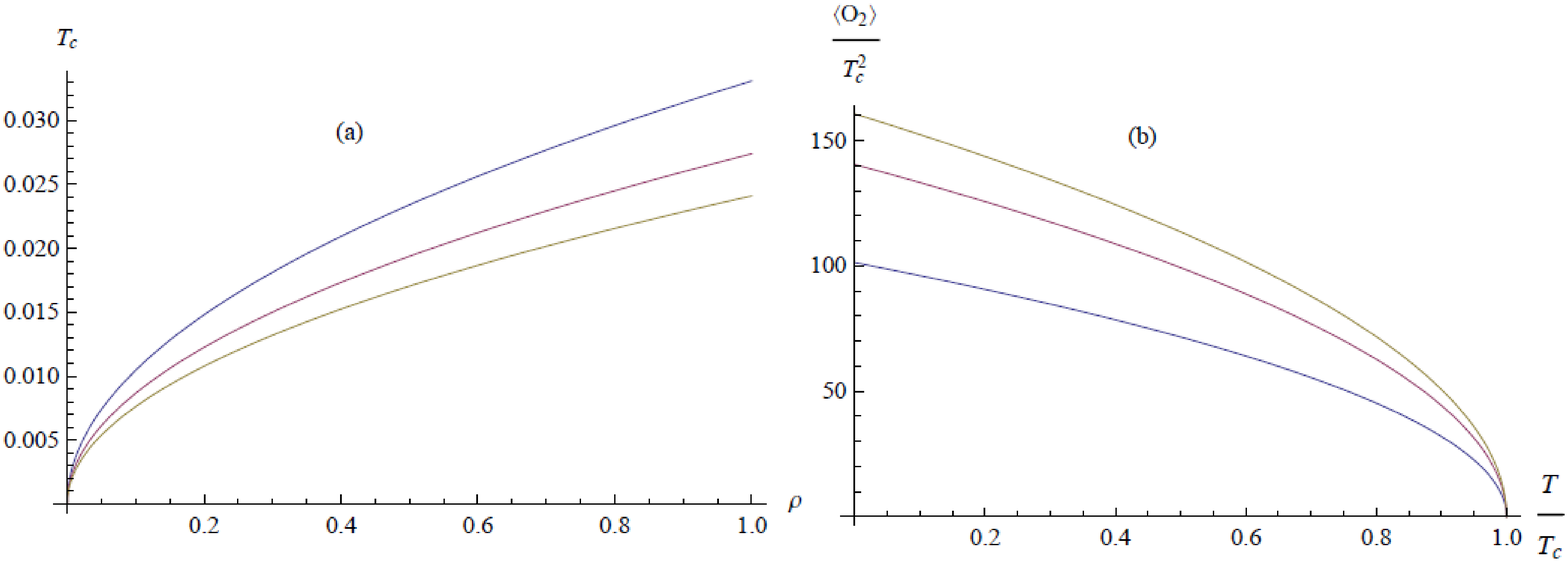}
\caption[]{\it (a)Critical temperature ($ T_{c}-\rho $) plot for $ s $-wave holographic superconductors (with $ z_m=1/2 $) for different choice of BI parameters $ b$. The upper curve corresponds to $ b=0 $, middle one corresponds to $ b=0.006 $ and the lower curve corresponds to $ b=0.009 $. (b) Condensation operator ($\frac{<\mathcal{O}_{2}>}{T_{c}^{2}}$) plot for various choice of BI parameter, $ b=0 $ (lower curve), $ b=0.006 $ (middle curve) and $ b=0.009 $ (upper curve).}
\label{figure 2a}
\end{figure}

Before we proceed further, let us make some comments on the results that have been obtained so far. First of all, from (\ref{neq21}) we note that, in order to have a meaningful notion for the critical temperature ($ T_c $) we must have the following upper bound 
\begin{equation}
b\leq \frac{1}{2\tilde{\beta}^{2}(1-z_m)} = \frac{1}{8(7-z_m)}
\end{equation} 
on the BI coupling parameter. Also we note that, the presence of above non linear (BI) corrections essentially lowers the critical temperature (see fig.1) which makes it harder to form the scalar hair at low temperature.

 
In order to study the effect of an external static magnetic field on a holographic superconductor we add a magnetic field in the bulk. According to the gauge gravity duality, the asymptotic value of this magnetic field corresponds to a magnetic field added to the boundary field theory, i.e, $ B({\bf{x}})=F_{xy}({\bf{x}},z\rightarrow 0) $ . Since the condensate is small everywhere near the upper critical value ($ B_{c} $) of the magnetic field, therefore we may regard the scalar field $ \psi $ as a perturbation near the critical field strength $ B\sim B_{c} $. Based on the above physical arguments, we adopt the following ansatz \cite{ref32},
\begin{equation}
A_t = \phi(z),~~~A_y = B x,~~~and~~~\psi = \psi (x,z)\label{neq24}.
\end{equation}

With the above choice, the scalar field equation for $ \psi $ turns out to be,
\begin{eqnarray}
\psi''(x,z)+\frac{f^{'}(z)}{f(z)}\psi'(x,z)+ \frac{r_+^{2}\phi^{2}(z)\psi(x,z)}{z^{4}f^{2}(z)}+\frac{2r_+^{2}\psi(x,z)}{z^{4}f(z)}+\frac{1}{z^{2}f(z)}(\partial_{x}^{2}\psi - B^{2}x^{2}\psi)=0\label{neq25}.
\end{eqnarray}
 
In order to solve the solve (\ref{neq25}), we take the following separable form
\begin{equation}
\psi (x,z) = X(x) R(z)\label{neq26}.
\end{equation}
 
 Substituting (\ref{neq26}) into (\ref{neq25}) we finally obtain,
 \begin{eqnarray}
 z^{2}f(z)\left[\frac{R''}{R}+ \frac{f'}{f}\frac{R'}{R}+\frac{\phi^{2}r_+^{2}}{z^{4}f^{2}}+\frac{2r_{+}^{2}}{z^{4}f} \right]-\left[ -\frac{X''}{X} + B^{2} x^{2}\right]=0.  
 \end{eqnarray}
 
 The equation for $ X(x)$ could be identified as the Schrodinger equation for a simple harmonic oscillator localized in one dimension with frequency determined by $ B $ \cite{ref32},
 \begin{equation}
 -X^{''}(x) + B^{2}x^{2}X(x) = \lambda_{n} B X(x)
 \end{equation}
 where $ \lambda_{n}= 2n+1 $ denotes the separation constant. We will be considering the lowest mode ($ n=0 $) solution which is expected to be most stable\cite{ref11,ref32}.  
 
 With this particular choice, the equation of $ R(z) $ turns out to be,
 \begin{equation}
 R''(z)+\frac{f^{'}(z)}{f(z)}R'(z)+ \frac{r_+^{2}\phi^{2}(z)R(z)}{z^{4}f^{2}(z)}+\frac{2r_+^{2}R(z)}{z^{4}f(z)}=\frac{B R(z)}{z^{2}f(z)}\label{neq27}.
 \end{equation}
 
 At the horizon ($ z=1 $), from (\ref{neq27}) one can obtain the following relation,
 \begin{equation}
 R'(1) = \left(\frac{2}{3} - \frac{B}{3r_+^{2}}\right) R(1)\label{neq28}. 
 \end{equation}
 On the other hand, the asymptotic solution ($ z\rightarrow 0 $) for (\ref{neq27}) may be written as,
 \begin{equation}
 R(z) = J_- z + J_+ z^{2} \label{neq29}
 \end{equation}
 where according to our previous choice $ J_- = 0 $.
 
 Near $ z=1 $, we may Taylor expand $ R(z) $ as, 
 \begin{equation}
 R(z) = R (1) - R^{'}(1)(1-z) + \frac{1}{2} R^{''}(1)(1-z)^{2} + .. ..\label{neq30}
 \end{equation}
 
 In order to calculate $ R'' (1) $ we will be using (\ref{neq27}). Considering (\ref{neq27}) close the horizon ($ z=1 $) we find,
 \begin{eqnarray}
 R''(1) &=& - \left[ \frac{f' R''+f'' R'}{f'(z)}\right]_{z=1}-\left[ \frac{r_+^{2}R(z)(-\phi'(1)(1-z)+..)^{2}}{(1-z)^{2}f'^{2}(z)}\right]_{z=1} -\frac{2r_+^{2}R'(1)}{f'(1)} +\frac{B R'(1)}{f'(1)}\nonumber\\
&=& -R''(1) - \frac{f''(1)}{f'(1)}R'(1)-\frac{r_+^{2}R(1)\phi'^{2}(1)}{f'^{2}(1)}-\frac{2r_+^{2}R'(1)}{f'(1)} +\frac{B R'(1)}{f'(1)}. 
 \end{eqnarray}
After some simple algebraic steps and using (\ref{neq28}) we finally obtain,
\begin{equation}
R''(1) = \left[\frac{8}{9}-\frac{5B}{9r_+^{2}}-\frac{\phi'^{2}(1)}{18 r_+^{2}} +\frac{B^{2}}{18r_+^{4}}\right]R(1)\label{neq31}. 
\end{equation} 
 Substituting (\ref{neq31}) into (\ref{neq30}) and using (\ref{neq28}) we find,
 \begin{equation}
 R(z) = \frac{1}{3}R(1) + \frac{2}{3}R(1)z+\frac{B R(1)}{3r_+^{2}}(1-z)+\frac{1}{2}\left[\frac{8}{9}-\frac{5B}{9r_+^{2}}-\frac{\phi'^{2}(1)}{18 r_+^{2}} +\frac{B^{2}}{18r_+^{4}}\right]R(1)(1-z)^{2}\label{neq32}. 
 \end{equation}

Using the previous technique, viz, matching the above two solutions (\ref{neq29},\ref{neq32}) for some intermediate point $ z=z_m $, we find the following set of equations,
\begin{equation}
J_+z_m^{2} = \frac{1}{3} R(1)+\frac{2}{3} R(1)z_m + \frac{B}{3r_+^{2}} R(1)(1-z_m)+ \frac{1}{2}\left[\frac{8}{9}-\frac{5B}{9r_+^{2}}-\frac{\phi'^{2}(1)}{18 r_+^{2}} +\frac{B^{2}}{18r_+^{4}}\right]R(1)(1-z_m)^{2}\label{neq33}
\end{equation}
and, 
\begin{equation}
2J_+z_m = \frac{2}{3}R(1) - \frac{B}{3r_+^{2}} R(1) - \left[\frac{8}{9}-\frac{5B}{9r_+^{2}}-\frac{\phi'^{2}(1)}{18 r_+^{2}} +\frac{B^{2}}{18r_+^{4}}\right]R(1)(1-z_m)\label{neq34}.
\end{equation}
 
 From the above set of equations (\ref{neq33},\ref{neq34}) it is quite trivial to find out the following quadratic equation in $ B $,
 \begin{equation}
 B^{2}+2Br_+^{2}\left( \frac{1+2z_m}{1-z_m}\right) +\frac{4r_+^{4}(7-z_m)}{(1-z_m)}-\phi'^{2}(1)r_+^{2} = 0
 \end{equation}
which has a solution,
\begin{equation}
B= \sqrt{\phi'^{2}(1)r_+^{2}-\frac{9r_+^{4}(3-4z_m)}{(1-z_m)^{2}}}-r_+^{2}\left(\frac{1+2z_m}{1-z_m} \right) \label{nbc}.
\end{equation} 
 
 Now consider the case for where the value of the external magnetic field ($ B $) is very close to the upper critical value i.e, $ B\sim B_{c} $. This implies a vanishingly small condensation and therefore we may ignore all the quadratic terms in $ \psi $. With this approximation from (\ref{neq5}) we obtain, 
 \begin{eqnarray}
\phi''(z)+\frac{2bz^3}{r_{+}^2 }\phi'^{3}(z) = 0\label{neq35}
\end{eqnarray}
which has a solution,
\begin{equation}
\phi (z) = \frac{\rho}{r_+}(1-z)\left(1-\frac{b\rho^{2}}{10r_+^{4}}\zeta(z) \right)+\mathcal{O}(b^{2})\label{neq36} 
\end{equation} 
with, $ \zeta(z) = 1+z+z^{2}+z^{3}+z^{4} $. 
 
Near the boundary ($ z\rightarrow 0 $) of the AdS space, we can approximately write (\ref{neq36}) as,
\begin{equation}
\phi (z) \simeq \frac{\rho}{r_+}\left(1-\frac{b\rho^{2}}{10r_+^{4}}\right) - \frac{\rho z}{r_+}\label{neq37}.
\end{equation} 
 Comparing (\ref{neq37}) with the first relation in (\ref{neq8}) one finds,
 \begin{equation}
 \mu = \frac{\rho}{r_+}\left(1-\frac{b\rho^{2}}{10r_+^{4}}\right)\label{eq38}.
 \end{equation}
 
 Near the horizon ($ z=1 $), from (\ref{neq35}) we obtain,
 \begin{equation}
 \phi''(1) = - \frac{2b}{r_+^{2}}\phi'^{3}(1)\label{eq39}.
\end{equation}  
 Substituting (\ref{eq39}) into (\ref{neq9}) we obtain,
 \begin{equation}
 \phi (z) = -\phi'(1)(1-z) - \frac{b}{r_+^{2}}\phi'^{3}(1)(1-z)^{2}\label{eq40}.
 \end{equation}
 
 Matching the solutions (\ref{neq8}) and (\ref{eq40}) at $ z=z_m $, we obtain the following set of equations
 \begin{equation}
 \mu - \frac{\rho z_m}{r_+} = -\phi'(1)(1-z_m) - \frac{b\phi'^{3}(1)}{r_+^{2}}(1-z_m)^{2}\label{eq41}
\end{equation}  
 and,
\begin{equation}
-\frac{\rho}{r_+} = \phi'(1)+ \frac{2b\phi'^{3}(1)}{r_+^{2}}(1-z_m)\label{eq42}.
\end{equation} 
 
 Finally, using (\ref{eq38}) the above set of equations (\ref{eq41},\ref{eq42}) may be written as,
 \begin{equation}
 \frac{\rho}{r_+}(1-z_m) + \phi'(1)(1-z_m) = \frac{b\rho^{3}}{10r_+^{5}}-\frac{b\phi'^{3}(1)}{r_+^{2}}(1-z_m)^{2}\label{eq43}
 \end{equation}
 and,
 \begin{equation}
 -\frac{\rho}{r_+} -\phi'(1) = \frac{2b\phi'^{3}(1)}{r_+^{2}}(1-z_m)\label{eq44}
 \end{equation}
 respectively.
Using the above equations (\ref{eq43},\ref{eq44}) the following relation could be easily obtained as,
\begin{equation}
\left(\beta^{4}+\beta^{3}\eta \right)(1-z_m)^{2} + \frac{1}{10} (\beta\eta^{3}+\eta^{4}) = 0 \label{eq45}
\end{equation}
where, $ \beta = \phi'(1) $ and $ \eta = \frac{\rho}{r_+} $. This is a quartic equation in $ \beta $.
One of the solutions for (\ref{eq45}) may be written as,
\begin{eqnarray}
\beta = -\eta ~~~~ \Rightarrow \phi'(1) = -\frac{\rho}{r_+}\label{neq46}.
\end{eqnarray}
 \begin{figure}[h]
\centering
\includegraphics[angle=0,width=16cm,keepaspectratio]{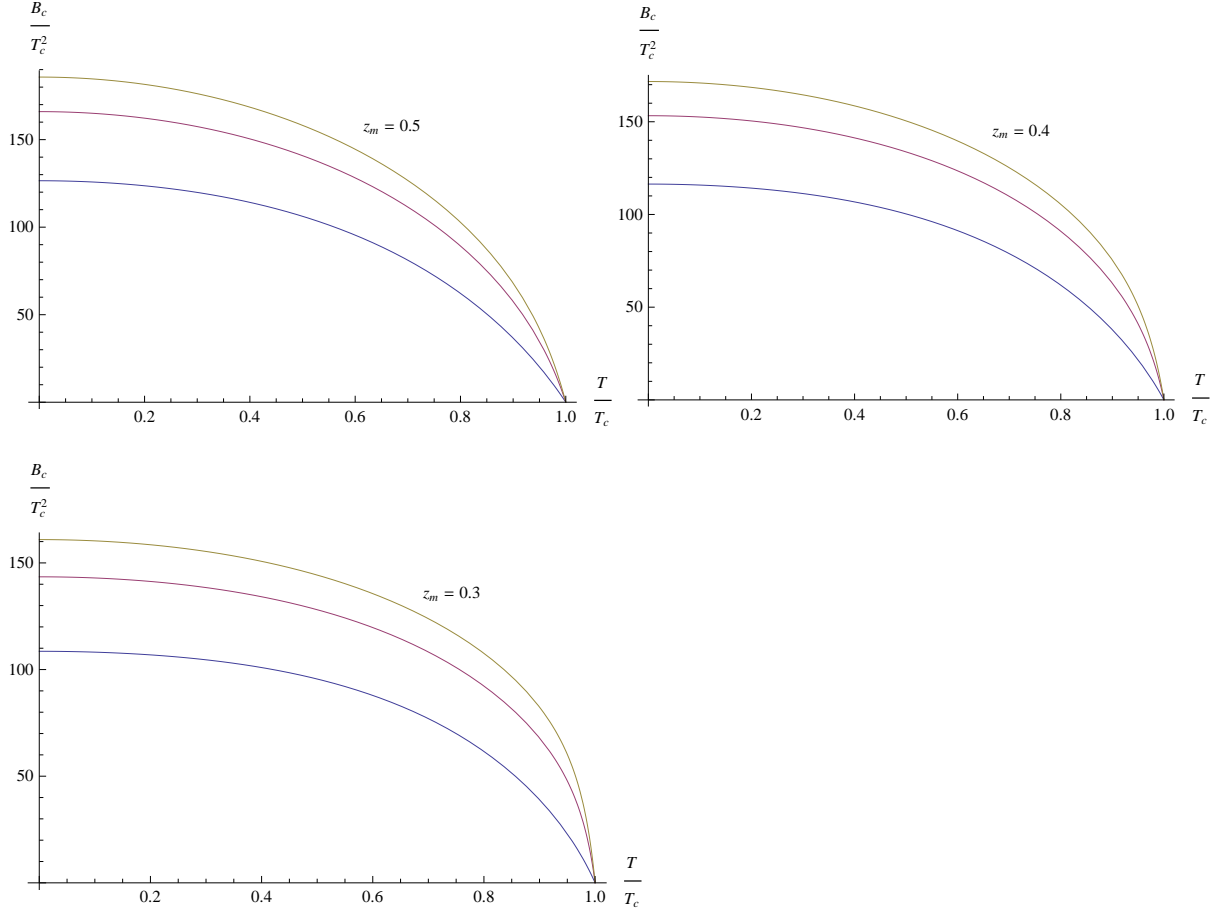}
\caption[]{\it Critical magnetic field ($ B_{c}$) plots for $ s $-wave holographic superconductors for different choice of BI parameters ($ b$), $ b=0 $ (lower curve), $ b=0.006 $ (middle curve) and $ b=0.009 $ (upper curve).}
\label{figure 2a}
\end{figure}

Substituting (\ref{neq46}) into (\ref{nbc}) and using (\ref{ntemp},\ref{neq21}), after a few algebraic steps we finally arrive at the following expression for the critical value of the magnetic field strength,

\begin{eqnarray}
B_{c} \simeq \frac{16\pi^{2}T_c^{2}(0)}{9}\left(1 + \frac{2 b\tilde{\beta}^{2}(1-z_m)}{1-\frac{9(3-4z_m)}{4(1-z_m)(7-z_m)}\left(\frac{T}{T_c} \right)^{4} }\right)\times \nonumber\\
\left[\tilde{\beta}\sqrt{1-\frac{9(3-4z_m)}{4(1-z_m)(7-z_m)}\left(\frac{T}{T_c} \right)^{4} } - \left(\frac{1+2z_m}{1-z_m} \right)  \left( \frac{T}{T_c}\right)^{2}\right] + \mathcal{O}(b^{2}).   
\end{eqnarray}   
 
In the above figure (2) the variation of the critical field strength has been depicted for different choices of the matching points ($ z_m=0.5,0.4, and ~~0.3 $). From these plots it is evident that the qualitative feature of individual plots does not alter due to different choices in the values of $ z_m $. The only quantitative change (for a given value of $ b $) that one can notice is that the value of the critical field strength is lowered for the smaller values of $ z_m $. Moreover, considering a particular case i.e; for a given $ z_m $, from these plots (fig.2) one can note that as the temperature of the $ s $-wave superconductor is lowered through the critical temperature $ T_c(0) $ (which is the critical temperature corresponding to zero magnetic field) the critical field strength gradually increases from zero to it's maximum value.  At this stage, one should note that the critical value ($ B_{c} $) of the magnetic field strength is higher in presence of non linear corrections than in the usual Maxwell case. This upper critical value increase as we increase the value of the BI coupling parameter ($ b $). The increasing field strength will try to reduce the condensate away completely, which implies that in the presence of higher derivative corrections to the usual Maxwell action the condensation gets harder to form.

 \section{Magnetic field effect in presence of Weyl corrections}
 
 In this section we aim to carry out our analysis in presence of Weyl corrections to the original Maxwell action. In order to do that we begin with the action for a ($ 3+1 $) Weyl corrected holographic superconductor \cite{nref4},
\begin{equation}
S= \int d^{5}x \sqrt{-g}\left[ \frac{1}{16 \pi G_N}\left( R + \frac{12}{L^{2}} \right) +  \mathcal{L}_{m}\right] 
\end{equation}
with,
\begin{equation}
\mathcal{L}_{m} = - \frac{1}{4}\left( F^{\mu\nu}F_{\mu\nu} - 4\gamma C^{\mu\nu\rho\sigma}F_{\mu\nu}F_{\rho\sigma}\right)- \frac{1}{L^{2}} |\nabla_{\mu}\psi -iA_{\mu}\psi |^{2} - \frac{m^{2}}{L^{4}}|\psi |^{2},
\end{equation}
where $ G_N $ is the gravitational Newton's constant, $ \frac{12}{L^{2}} $ corresponds to a negative cosmological constant, $ \gamma $ ($-\frac{L^{2}}{16} < \gamma < \frac{L^{2}}{24} $) is a dimensionful constant and $ C_{\mu\nu\rho\sigma} $ is the Weyl tensor. In the following analysis, without loss of any generality we set $ L = 1 $.

The metric for the planar Schwarzschild AdS black hole may be written as,
\begin{equation}
ds_{5}^{2} = r^{2}(-f(r)dt^{2} + dx_i dx^{i}) + \frac{dr^{2}}{r^{2}f(r)}
\end{equation}
with,
\begin{equation}
f(r) = 1 - \frac{r_+^{4}}{r^{4}}
\end{equation}
where $ r_+ $ is the radius of the outer event horizon.

The Hawking temperature of the black hole may be written as,
\begin{equation}
T = \frac{r_+}{\pi}\label{temp}.
\end{equation}

Equations of motion for the Maxwell field ($ A_{\mu} $) and the complex scalar field ($ \psi $) may be written as,
\begin{equation}
\frac{1}{\sqrt{-g}} \partial_{\mu}(\sqrt{-g}F^{\mu\nu}) - \frac{4 \gamma}{\sqrt{-g}} \partial_{\mu} (\sqrt{-g}C^{\mu\nu\rho\sigma} F_{\rho\sigma})-i\left( \psi^{*}\partial^{\nu}\psi - \psi(\partial^{\nu}\psi)^{*}\right)-A^{\nu} |\psi |^{2} = 0 \label{eq1}
\end{equation} 
and,
\begin{equation}
\partial_{\mu}\left(\sqrt{-g}\partial^{\mu}\psi \right) -i\sqrt{-g}A^{\mu}\partial_{\mu}\psi -i\partial_{\mu}\left(\sqrt{-g}A^{\mu}\psi \right)-\sqrt{-g}A^{2}\psi -\sqrt{-g}m^{2}\psi = 0 \label{eq2}.
\end{equation}

Considering the following ansatz,
\begin{eqnarray}
A_{\mu}=(\phi(r),0,0,0,0),\;\;\;\;\psi=\psi(r)
\label{vector}
\end{eqnarray}
the above set of equations (\ref{eq1},\ref{eq2}) turns out to be,
\begin{equation}
\left(1 - \frac{24 \gamma r_+^{4}}{r^{4}} \right) \phi''(r) + \left( \frac{3}{r} + \frac{24 \gamma r_+^{4}}{r^{5}}\right) \phi'(r) - \frac{2 \phi \psi^{2}}{r^{2}f} = 0 \label{eq3}  
\end{equation}
and,
\begin{equation}
\psi''(r) + \left( \frac{5}{r} + \frac{f^{'}}{f}\right) \psi'(r) + \frac{\phi^{2}\psi}{r^{4}f^{2}} - \frac{m^{2}\psi}{r^{2}f} = 0\label{eq4} 
\end{equation}
respectively.

In order to express the above set of equations in a more suitable way, we set $ m^{2} = -3 $ and choose $ z=\frac{r_+}{r} $. With this choice of variable, (\ref{eq3}) and (\ref{eq4}) turn out to be,
\begin{equation}
\phi''(z) - \frac{1}{z} \left(\frac{1 + 72 \gamma z^{4}}{1 - 24 \gamma z^{4}} \right) \phi'(z) - \frac{2\psi^{2}(z)\phi(z)}{z^{2}(1 - 24 \gamma z^{4})f(z)} = 0 \label{eq5}
\end{equation}
and,
\begin{equation}
\psi''(z) + \left( \frac{f'(z)}{f(z)} - \frac{3}{z}\right)\psi'(z) + \frac{\phi^{2}(z)\psi(z)}{r_+^{2} f^{2}(z)} + \frac{3\psi (z)}{z^{2}f(z)} = 0 \label{eq6}
\end{equation}
respectively.

Before going further, it is customary to mention about the boundary conditions.
\vskip 1mm
\noindent
$\bullet$ Boundary conditions at the horizon ($ z=1 $) may be written as,
\begin{equation}
\phi (1) = 0,~~~ ~~~\psi'(1) = \frac{3}{4} \psi (1)\label{eq7}.
\end{equation}
\vskip 1mm
\noindent
$\bullet$ The asymptotic ($ z\rightarrow 0 $) boundary conditions may be expressed as,
\begin{eqnarray}
\phi (z) = \mu - \frac{\rho z^{2}}{r_+^{2}}\label{eq8}\\
\psi (z) = J_- z +  J_+ z^{3} \label{eq9}
\end{eqnarray}  
where $ \mu $ and $ \rho $ are the chemical potential and the charge density of the boundary field theory. On the other hand $ J_+ $ is related to the vacuum expectation of the condensation operator ($ <\mathcal{O}_+> $) in the dual field theory and $ J_- $ acts as a source. Since we want condensation with out being sourced, therefore we set $ J_- = 0 $.

  In order to investigate the effect of an external magnetic field on $ s $ wave condensation, as a first step, it is essential to compute the critical temperature ($ T_c(0) $) in the absence of external magnetic field. To do that, we first Taylor expand $ \phi (z) $
and $ \psi (z) $ near the horizon ($ z = 1 $) as, 
\begin{eqnarray}
\phi (z)=\phi (1) - \phi^{'}(1)(1-z) + \frac{1}{2}\phi^{''}(1)(1-z)^{2} + .. ..\label{eq10}
\end{eqnarray}
and,
\begin{eqnarray}
\psi(z)=\psi(1)-\psi^{'}(1)(1-z)+\frac{1}{2} \psi^{''}(1)(1-z)^{2} + ....\label{eq11}
\end{eqnarray}
respectively, where without loss of generality we choose $ \phi'(1)<0 $ and $ \psi(1)>0 $.

From (\ref{eq5}) we note that,
\begin{equation}
\phi''(1) = \frac{\phi'(1)}{1-24 \gamma}\left[1 + 72 \gamma - \frac{\psi^{2}(1)}{2} \right]\label{eq12}. 
\end{equation}
Substituting (\ref{eq12}) into (\ref{eq10}) we finally obtain,
\begin{equation}
\phi (z) = -\phi'(1) (1-z) + \frac{\phi'(1)}{2(1-24 \gamma)}\left[1 + 72 \gamma - \frac{\psi^{2}(1)}{2} \right] (1-z)^{2}\label{eq13}.
\end{equation}

On the other hand, from (\ref{eq6}) and using (\ref{eq7}) we obtain,
\begin{equation}
\psi'' (1) = \frac{9}{32} \psi (1) - \frac{\phi'^{2}(1)\psi (1)}{32 r_+^{2}}\label{eq14}.
\end{equation}
Finally, substituting (\ref{eq14}) into (\ref{eq11}) we find,
\begin{equation}
\psi (z) = \psi (1) - \frac{3}{4} \psi (1) (1-z) + \frac{1}{2} \left(\frac{9}{32} -  \frac{\phi'^{2}(1)}{32 r_+^{2}} \right) \psi (1) (1-z)^{2}\label{eq15}.
\end{equation}

As a next step, following the matching method \cite{ref19}, we match the solutions (\ref{eq8},\ref{eq13}) and (\ref{eq9},\ref{eq15}) at some intermediate point $ z = z_m $, which yields the following set of equations,
\begin{eqnarray}
\mu - \frac{\rho z_m^{2}}{ r_+^{2}} &=& \beta (1-z_m) - \frac{\beta}{2(1-24\gamma)}\left[1 + 72\gamma - \frac{\alpha^{2}}{2} \right](1-z_m)^{2}\\
- \frac{2\rho z_m}{r_+^{2}} &=& -\beta + \frac{\beta(1 + 72\gamma)}{(1-24\gamma)}\left[1 - \frac{\alpha^{2}}{2(1 + 72\gamma)} \right](1-z_m)\label{eq16}\\
J_+z_m^{3} &= &\alpha - \frac{3}{4}\alpha (1-z_m) + \frac{\alpha}{2} \left(\frac{9}{32} - \frac{\tilde{\beta}^{2}}{32}\right)(1-z_m)^{2}\label{eq17}\\
J_+z_m^{2} &=& \frac{\alpha}{4} - \frac{\alpha}{3} \left(\frac{9}{32} - \frac{\tilde{\beta}^{2}}{32}\right)(1-z_m)\label{eq18}  
\end{eqnarray}
where, $ \beta = -\phi'(1) $, $ \psi (1) = \alpha $ and $ \tilde{\beta} = \beta/r_+ $.

After some algebraic steps, from (\ref{eq16}) we obtain,
\begin{equation}
\alpha^{2} =  \frac{4 z_m \rho (1 - 24\gamma)}{\pi^{3} T^{3} \tilde{\beta}(1-z_m)} \left(1 - \frac{T^{3}}{T_c^{3}} \right)\label{eq19} 
\end{equation}
with the identification of the critical temperature 
\begin{equation}
T_c = \left[\frac{2 z_m}{\pi^{3} \tilde{\beta}}\left( \frac{1 - 24\gamma}{z_m - 24\gamma(4-3z_m)}\right) \right]^{\frac{1}{3}} \rho^{1/3}\label{eq20}
\end{equation}
where,  $ \gamma < \frac{z_m}{24 (4-3 z_m)} $.

From (\ref{eq17}) and (\ref{eq18}) it is easy to show that,
\begin{equation}
\tilde{\beta} = \frac{\sqrt{9z_m^{2}+60z_m+75}}{\sqrt{(1-z_m)(3-z_m)}}.
\end{equation}

On the other hand, using (\ref{eq17}), (\ref{eq18}), (\ref{eq19}) and (\ref{eq20}) one can easily find,
\begin{equation}
J_+ = \frac{(3z_m + 5)}{4z_m^{2}(3-z_m)} \alpha =\frac{(3z_m + 5)}{4z_m^{2}(3-z_m)}\sqrt{\frac{2}{1-z_m}} \sqrt{z_m- 24\gamma(4-3z_m)} \left( \frac{T_c}{T}\right)^{\frac{3}{2}}\sqrt{1 - \frac{T^{3}}{T_c^{3}}} \label{eq21}. 
\end{equation}   

Finally, using (\ref{temp}) the condensation operator for $ T\sim T_c $ may be found as,
\begin{equation}
<\mathcal{O}_+> = J_+ r_+^{3} \simeq \frac{\sqrt{3}(3z_m + 5)\pi^{3}T_c^{3}}{4z_m^{2}(3-z_m)}\sqrt{\frac{2}{1-z_m}} \sqrt{z_m- 24\gamma(4-3z_m)} \sqrt{1 - \frac{T}{T_c}}.
\end{equation} 



With the above expressions in hand, we are now in a position to study the effect of an external static magnetic field on a holographic superconductor in presence Weyl corrections. In order to study the consequences of applying an external magnetic field ($ B $) in the dual field theory we adopt the following ansatz \cite{ref15}, \cite{ref32},\cite{ref34},
\begin{equation}
A_t = \phi (z),~~~A_y = B x ~~~ and~~~ \psi = \psi (x,z).
\end{equation}

With the above choice, the equation for $ \psi (x,z) $ turns out to be,
\begin{equation}
\psi''(x,z) + \left(\frac{f'}{f} - \frac{3}{z} \right) \psi'(x,z) + \frac{\phi^{2}(z) \psi (x,z)}{r_+^{2} f^{2}(z)} + \frac{3 \psi (x,z)}{z^{2} f(z)} + \frac{1}{r_+^{2}f(z)} (\partial_x^{2} \psi - B^{2}x^{2}\psi) = 0 \label{eq21}. 
\end{equation}
In order to solve the above equation (\ref{eq21}) we take the solution of the following form,
\begin{equation}
\psi (x,z) = X(x) R(z).\label{eq22}
\end{equation}
Substituting (\ref{eq22}) into (\ref{eq21}) we find,
\begin{eqnarray}
r_+^{2}f(z)\left[ \frac{R''(z)}{R(z)} + \left(\frac{f'}{f} - \frac{3}{z} \right) \frac{R'(z)}{R(z)} + \frac{\phi^{2}(z)}{r_+^{2} f^{2}(z)} + \frac{3}{z^{2}f(z)}\right]- \left[-\frac{X''(x)}{X(x)} + B^{2}x^{2} \right]= 0. 
\end{eqnarray}
It is now straightforward to show that the equation corresponding to $ X(x) $ basically represents the motion of a one dimensional harmonic oscillator with frequency determined by $ B $ \cite{ref32}.
\begin{equation}
- X'' (x) + B^{2} x^{2} X(x) = \lambda_n B X(x)
\end{equation} 
 where $ \lambda_n = 2n+1 $ is the separation constant. In the following analysis we shall set $ n =0 $ thereby paying attention to the lowest mode of solutions.
 
 With this particular choice, the equation for $ R(z) $ turns out to be
 \begin{equation}
 R''(z) + \left(\frac{f'}{f} - \frac{3}{z} \right) R'(z) + \frac{\phi^{2} R(z)}{r_+^{2}f^{2}(z)} + \frac{3 R (z)}{z^{2} f(z)} = \frac{B R(z)}{r_+^{2} f(z)}\label{eq23}.
 \end{equation}
 
 Before we proceed further, let us first discuss the boundary conditions for (\ref{eq23}).
\vskip 1mm
\noindent
$\bullet$   
For $ z = 1 $ one can easily obtain,
\begin{equation}
R'(1) = \left( \frac{3}{4} - \frac{B}{4r_+^{2}}\right) R(1)\label{r}.
\end{equation}
\vskip 1mm
\noindent
$\bullet$ The asymptotic ($ z\rightarrow 0 $) behavior of (\ref{eq23}) may be written as
\begin{equation}
R (z) = J_- z + J_+ z^{3}\label{1}
\end{equation} 
with $ J_- = 0 $.

On the other hand, following our previous approach, we Taylor expand $ R(z) $ close to the horizon as,
\begin{equation}
R(z) = R(1) - R'(1) (1-z) + \frac{1}{2} R''(1) (1-z)^{2} + .. \label{eq24}
\end{equation} 
In order to compute $ R''(1) $ let us consider (\ref{eq23}) close to the horizon ($ z = 1 $), which yields,
\begin{eqnarray}
R''(1)&=&-\left[ \frac{f'' R'+f' R''}{f'}\right]_{z=1}+3R'(1)-\left[ \frac{(-\phi'(1)(1-z)+..)^{2} R(z)}{r_+^{2}(-f'(1)(1-z))^{2}}\right]_{z=1}-\frac{3R'(1)}{f'(1)}+\frac{B R'(1)}{r_+^{2}f'(1)}\nonumber\\
&=& \left[9 - \frac{6B}{r_+^{2}} + \frac{B^{2}}{r_+^{4}} - \frac{\phi'^{2}(1)}{r_+^{2}} \right]\frac{R(1)}{32}\label{eq25}. 
\end{eqnarray}
Substituting (\ref{eq25}) into (\ref{eq24}) and using (\ref{r})  we get,
\begin{equation}
R (z) = R(1) - \left[\frac{3}{4} - \frac{B}{4r_+^{2}} \right] R(1) (1-z) + \left[9 - \frac{6B}{r_+^{2}} + \frac{B^{2}}{r_+^{4}} - \frac{\phi'^{2}(1)}{r_+^{2}} \right]\frac{R(1)}{64}(1-z)^{2}\label{eq26} 
\end{equation}

Following the basic arguments of matching method \cite{ref19}, we match the equations (\ref{1}) and (\ref{eq26}) at some intermediate point $ z = z_m $ which finally yields the following set of equations,  
\begin{eqnarray}
J_+z_m^{3} = R(1) - \left[\frac{3}{4} - \frac{B}{4r_+^{2}} \right] R(1) (1-z_m) + \left[9 - \frac{6B}{r_+^{2}} + \frac{B^{2}}{r_+^{4}} - \frac{\phi'^{2}(1)}{r_+^{2}} \right]\frac{R(1)}{64}(1-z_m)^{2}\label{eq27}
\end{eqnarray}
and,
\begin{equation}
3z_m^{2}J_+ = \left[\frac{3}{4} - \frac{B}{4r_+^{2}} \right] R(1) -\left[9 - \frac{6B}{r_+^{2}} + \frac{B^{2}}{r_+^{4}} - \frac{\phi'^{2}(1)}{r_+^{2}} \right]\frac{R(1)}{32}(1-z_m) \label{eq28}.
\end{equation}

Using (\ref{eq27}) and (\ref{eq28}) it is now straightforward to obtain the following quadratic equation in $ B $, 
\begin{equation}
\zeta_1 B^{2} + \zeta_2 Br_+^{2} - \zeta_3 \phi'^{2}(1)r_+^{2} + \zeta_4 r_+^{4} = 0\label{eq29}
\end{equation}
where,
\begin{eqnarray}
\zeta_1 &=& (1-z_m)(3-z_m)=\zeta_3 \nonumber\\
\zeta_2 &=& 30-8z_m-6z_m^{2}\nonumber\\
\zeta_4 &=& 75+60z_m+9z_m^{2}
\end{eqnarray}
Solution of (\ref{eq29}) may be written as,
\begin{equation}
B = \frac{1}{2\zeta_1}\left[\sqrt{(\zeta_2^{2}-4\zeta_1 \zeta_4)r_+^{4} + 4\zeta_1\zeta_3\phi'^{2}(1)r_+^{2}} - \zeta_2 r_+^{2} \right]\label{b}. 
\end{equation}

Considering $ B\sim B_{c} $, we may ignore all the quadratic and other higher order terms in $ \psi $ as the condensation is very small near the critical field strength. With this argument (\ref{eq5}) becomes,
\begin{equation}
\phi''(z) - \frac{1}{z} \left(\frac{1 + 72 \gamma z^{4}}{1 - 24 \gamma z^{4}} \right) \phi'(z) = 0\label{eq30}.
\end{equation}
It is now trivial to note that for $ z\rightarrow 0 $ the solution for (\ref{eq30}) may be written as,
\begin{equation}
\phi (z) = \frac{\rho}{r_+^{2}} (1- z^{2})\label{eq31}
\end{equation}
where comparing with (\ref{eq8}) one can identify the chemical potential as $ \mu = \frac{\rho}{r_+^{2}} $.

As a next step, we consider (\ref{eq30}) close to the horizon ($ z = 1 $) which yields,
\begin{equation}
\phi''(1) = \left(\frac{1 + 72 \gamma}{1 - 24 \gamma} \right) \phi'(1)\label{eq32}.
\end{equation}
 Substituting (\ref{eq32}) into (\ref{eq10}) and then matching with (\ref{eq31}) for some intermediate value of $ z=z_m $ we eventually find the following set of equations,
\begin{figure}[h]
\centering
\includegraphics[angle=0,width=16cm,keepaspectratio]{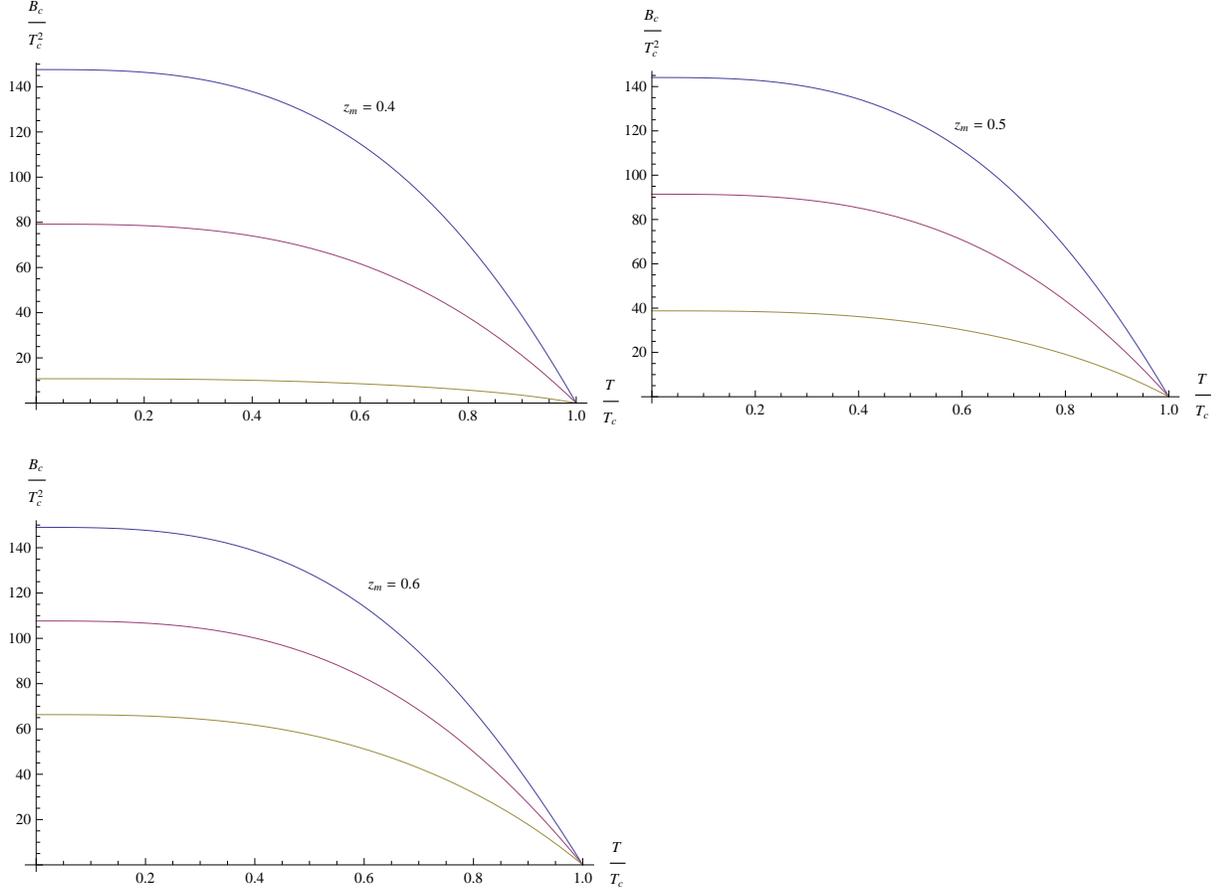}
\caption[]{\it Critical magnetic field strength ($ B_{c}-\frac{T}{T_c} $) plot for $ s $-wave holographic superconductors for different choice of Weyl coupling parameters $ \gamma $. The upper curve corresponds to $ \gamma=-0.006 $, middle curve corresponds to $ \gamma = 0 $ and the lower curve corresponds to $ \gamma=0.006 $.}
\label{figure 2a}
\end{figure}

\begin{equation}
\frac{\rho}{r_+^{2}} (1-z_m^{2}) = -\phi'(1) (1-z_m) + \frac{1}{2} \phi'(1) \left( \frac{1+72\gamma}{1-24\gamma}\right) (1-z)^{2}\label{eq34}
\end{equation}
and,
\begin{equation}
-\frac{2\rho z_m}{r_+^{2}} = \phi'(1) - \phi'(1)\left( \frac{1+72\gamma}{1-24\gamma}\right)(1-z_m)\label{eq35}.
\end{equation}
Finally, using (\ref{eq34},\ref{eq35}) it is quite trivial to show that,
\begin{equation}
\phi'(1) = -\frac{2\rho}{r_+^{2}}~~~\Rightarrow \phi'^{2}(1)r_+^{2} = \frac{4\rho^{2}}{r_+^{2}}\label{eq36}.
\end{equation}
Substituting (\ref{eq36}) into (\ref{b}) and using (\ref{temp}, \ref{eq20}) we finally obtain 
\begin{equation}
B_{c} \simeq \pi^{2} \tilde{\beta}T_c^{2}(0) \left(1-\frac{96\gamma (1-z_m)}{z_m\left( 1+\frac{(\zeta_2^{2}-4\zeta_1\zeta_4)}{4\tilde{\beta}^{2}\zeta_1\zeta_3}\left(\frac{T}{T_c} \right)^{6}\right) } \right)\left[ {\sqrt{1+\frac{(\zeta_2^{2}-4\zeta_1\zeta_4)}{4\tilde{\beta}^{2}\zeta_1\zeta_3}\left(\frac{T}{T_c} \right)^{6}}} - \frac{\zeta_2}{2\tilde{\beta}\sqrt{\zeta_1\zeta_3}}\left(\frac{T}{T_c} \right)^{3} \right] + \mathcal{O}(\gamma^{2})\label{bc}.
\end{equation}

From the above figure (3) one can note that the critical field strength ($ B_c $) increases as the temperature is lowered through ($ T_c $). This resembles the basic qualitative features of type II superconductors where the upper critical field strength ($ B_{c2} $) shows identical behavior at low temperatures\cite{poole}. Furthermore, from (\ref{bc}) we note that for $ T \sim T_c $
\begin{equation}
B_c \varpropto \left(1 - \frac{T}{T_c} \right) 
\end{equation}
which agrees well with the standard expression for the upper critical field strength as predicted by Ginzburg- Landau theory \cite{ref14},\cite{poole}. Moreover, for a given value of $ z_m $, the value of the critical field strength ($ B_{c} $) has been found to be increasing for $ \gamma<0 $, which indicates the onset of a harder condensation. One may also note that (for a given value of $ \gamma $) the critical field strength is higher for higher values of $ z_m $.

\section{Conclusions}
 
 In this paper, based on the matching technique \cite{ref19}, various properties of holographic $ s $- wave superconductors have been investigated in the the probe limit. The present paper aims to investigate the behavior of holographic superconductors immersed in an external magnetic field in the presence of various non linear corrections to the usual Maxwell action. These non linear corrections basically correspond to higher derivative corrections of the gauge fields. In the present work two such non linear corrections have been considered, namely (1) Born-Infeld (BI) correction and (2) Weyl correction. In both the cases it has been observed (upto \textit{leading} order in the coupling parameters ) that the properties of holographic superconductors are indeed affected due to the presence of these non linear effects.
The observations may be put as follows:
 \vskip 1mm
\noindent
$\bullet$  In both the examples, it has been observed that the critical temperature ($ T_c $) indeed gets affected due to the presence of these higher derivative corrections. In the first example, $ T_c $ has been found to be decreasing as we increase the strength of the BI coupling parameter ($b$), whereas in the second case it decreases for $ \gamma<0 $. These indicate that the condensation gets harder due to these non linear effects.
\vskip 1mm
\noindent
$\bullet$  In both the cases an upper bound in these coupling parameters have been found above which the analysis can not be carried out.
\vskip 1mm
\noindent
$\bullet$ In both the cases the critical exponent associated with the condensation value near the critical point has been found to be equal to $ 1/2 $ which is in good agreement with the universal mean field value and indeed suggests the onset of a second order phase transition near the critical temperature.  
\vskip 1mm
\noindent
$\bullet$ Finally, and most importantly it has been observed that the $ s $-wave condensate exists below certain critical value ($ B_c $) of the external magnetic field. In the first example it is observed that the value of the critical field strength ($ B_{c} $) increases as we increase the strength of the BI coupling  parameter ($ b $), whereas in the second example it is found that the critical field strength increases for $ \gamma<0 $. These in fact suggest that as the magnetic field strength increases it will try to reduce the condensate away completely and thereby making it harder to form the scalar hair at low temperatures. Moreover one can identify the critical field strength to that with the upper critical value of the magnetic field in type II superconductors which also satisfies the standard relation predicted by Ginzburg-Landau theory.

Finally, I would like to conclude the paper mentioning some of its future prospects. First of all, one can repeat the above calculations incorporating the back reaction on the metric. Although the above analysis has been carried out replacing the Maxwell action by a BI one, the above analysis may also be carried out for other non linear theories also, for example, replacing the Maxwell action by a power Maxwell action. It will be also interesting to generalize the above calculations in higher dimensions and extending the methodology for the Gauss Bonnet gravity in the framework of non linear electrodynamics.

\vskip 5mm
{\bf{Acknowledgement :}}\\
     Author would like to thank the Council of Scientific and Industrial Research (C. S. I. R), Government of India, for financial help.


\end{document}